\documentstyle[12pt]{article}
\input{epsf}
\newcommand{\be}{\begin{eqnarray}}
\newcommand{\ee}{\end{eqnarray}}
\newcommand{\ba}{\begin{array}}
\newcommand{\ea}{\end{array}}

\begin{document}
\renewcommand{\thefootnote}{\fnsymbol{footnote}}
%
%

\rightline{\textsl{}} \vspace{0.5cm}
\begin{center}
{\Large Tomography for amplitudes of hard exclusive processes}\\
\vspace{0.35cm}
 M.V. Polyakov\\

\vspace{0.35cm}
Petersburg Nuclear Physics
Institute, Gatchina, St.\ Petersburg 188350, Russia\\
and\\
Institut f\"ur Theoretische Physik II,
Ruhr--Universit\"at Bochum, D--44780 Bochum, Germany

%
%
\end{center}

\begin{abstract}
We discuss which part of information about hadron structure
encoded in the Generalized Parton Distributions (GPDs) [part of total GPD image]
can be restored from the known amplitude
of a hard exclusive process. The physics content of this partial image is analyzed.
Among other things, we show that this partial image contains direct information about how
the target hadron responses to the
(string) quark-antiquark operator of {\it arbitrary} spin $J$.
Explicit equations relating physics content of the partial image of GPDs directly to the data are derived.
Also some new results concerning the dual parametrization of GPDs are presented.

\end{abstract}
\vspace{0.1cm}

\section*{\normalsize \bf Introduction}
\noindent
Measurements of hard exclusive processes can provide us with rich information about hadron
structure encoded in generalized parton distributions (GPDs) \cite{pioneers}
(for recent reviews of GPDs see Refs.~\cite{GPV,Diehlrev,Belitskyrev}).
Frequently one speaks about
``hadron tomography", ``3D images of proton", or ``fempto-hologram", etc. Indeed, if we would know GPD as a functions
of all its variable we would determine, among many other things, the distributions of hadron's constituents (quarks and gluons) in three dimensional
space. The amplitudes of hard exclusive processes, which we consider as direct observables, are given
by the convolution of GPDs with perturbative kernels. It means that measurements of the amplitudes
provide us with some kind of sectional  images of GPDs -- the convolution integral ``projects out" one
of variables in GPDs. This is a typical problem of tomography which is usually solved with help of Radon transformation
\cite{Radon}. Recently the Radon transformation was applied in Ref.~\cite{Teryaev:2001qm} in order to
restore the double distributions from known GPDs.

In the present paper we address the question:
What part of the hadron image provided by GPDs can be reconstructed
if we know the amplitude of hard exclusive process? We give explicit formulae
which relate the forward-like
parton distributions of dual parametrization of GPDs \cite{MaxAndrei} to the amplitude of hard exclusive processes.
We shall see that at fixed renormalization scale it is impossible to restore the
complete GPD image, however, part of the total image carries valuable information about hadron structure.
Furthermore, studying of the scaling violation in hard exclusive processes would allow, in principle, to restore total GPD.
We suggest a practical way to determine GPDs from data on hard exclusive processes.

We restrict ourselves to hard exclusive processes at the leading order (LO) of perturbative expansion
as this order provides the bulk information about the hadron structure. The perturbative corrections can
be included into our analysis.
Recently similar problems, from other point of view,
were addressed in Refs.~\cite{Mueller:2006pm,Kumericki:2007sa} including the NLO correction.

\section*{\normalsize \bf Basics of the dual parametrization of GPDs}
\noindent
For our analysis we employ dual parametrization of GPDs suggested in Ref.~\cite{MaxAndrei}. This parametrization
is based on representation of parton
distributions as an infinite series of $t$-channel exchanges \cite{MVP98}.
The dual parametrization has been already used to describe the data on deeply virtual Compton scattering
in Ref.~\cite{Guzey}. In this section we remind the basics of the dual parametrization and present also some
new results.

In the dual parametrization GPDs $H(x,\xi,t)$
is expressed in terms of set of functions $Q_{2\nu}(x,t)$ ($\nu=0,1,2,\ldots$) as \cite{MaxAndrei}:

\be
\label{integralrelation}
H(x,\xi,t)&=&\sum_{\nu=0}^{\infty}\Biggl\{ \frac{\xi^{2\nu}}{2}\
\left[ H^{(\nu)}(x,\xi,t)-H^{(\nu)}(-x,\xi,t)\right]\\
\nonumber
&-&
\left(1-\frac{x^2}{\xi^2}\right)\theta(\xi-|x|)\sum_{l=1\atop \scriptstyle{\rm odd}}^{2\nu-3}
C^{3/2}_{2\nu-l-2}\left(\frac{x}{\xi}\right)\ P_l\left(\frac
1\xi\right)\int_{0}^1 dy\ y^{2\nu-l-2}\ Q_{2\nu}(y,t)\Biggr\}\, ,
\ee
where the function $H^{(\nu)}(x,\xi,t)$ is defined on the interval $-\xi \leq x\leq 1$ and it is given as the following
integral transformation:

\be
\nonumber
H^{(\nu)}(x,\xi,t)&=&\theta\left(x>\xi\right)\ \frac{1}{\pi}
\int_{y_0}^1 \frac{dy}{y}\left[\left(1-y\frac{\partial}{\partial y}\right)
Q_{2\nu}(y,t)\right]
\int_{s_1}^{s_2} ds \frac{x_s^{1-2\nu}}{\sqrt{2x_s-x_s^2-\xi^2}}\\
&+&
\label{kernel-podrobno}
\theta\left(x<\xi\right)\ \frac{1}{\pi}
\int_{0}^1 \frac{dy}{y}\left[\left(1-y\frac{\partial}{\partial y}\right)
Q_{2\nu}(y,t)\right]
\int_{s_1}^{s_3} ds \frac{x_s^{1-2\nu}}{\sqrt{2x_s-x_s^2-\xi^2}}\, .
\ee
Here $x_s=2\frac{x-s\xi}{(1-s^2)y}$ and integration limits $s_1,s_2, s_3$ and $y_0$ are
given by the following expressions:

\be
\nonumber
s_1&=&\frac{1}{y\xi}\left[1-\sqrt{1-\xi^2}-\sqrt{2\left(1-x
y\right)\left(1-\sqrt{1-\xi^2}\right)-\xi^2\left(1-y^2\right)}\ \right]\, ,\\
\nonumber
s_2&=&\frac{1}{y\xi}\left[1-\sqrt{1-\xi^2}+\sqrt{2\left(1-x
y\right)\left(1-\sqrt{1-\xi^2}\right)-\xi^2\left(1-y^2\right)}\ \right]\, ,\\
\nonumber
s_3&=&\frac{1}{y\xi}\left[1+\sqrt{1-\xi^2}-\sqrt{2\left(1-x
y\right)\left(1+\sqrt{1-\xi^2}\right)-\xi^2\left(1-y^2\right)}\ \right]\, ,\\
y_0&=&\frac{1}{\xi^2}\left[x \left(1-\sqrt{1-\xi^2}\right)+
\sqrt{ \left(x^2-\xi^2\right)\left(2\left(1-\sqrt{1-\xi^2}\right)-\xi^2\right)}\ \right]\, .
\ee
 The resulting GPD automatically satisfy
polynominality property, all sum rules and have correct scale dependence.

Let us remind the main properties
of the forward-like functions $Q_{2\nu}(x,t)$
\footnote{We restrict ourselves to the case the spinless hadron and singlet distributions. Generalization
for other cases is trivial}:

\begin{itemize}
\item
At the LO scale dependence of functions $Q_{2\nu}(x,t)$ is given by the standard DGLAP evolution equation, so that these functions
behave as usual parton distributions under QCD evolution.

\item
The function $Q_0(x,t)$ at $t=0$ is related to the forward distribution $q(x)$ as:

\be
Q_0(x,t=0)=\left[ (q(x)+\bar q(x))-\frac x2  \int_x^1 \frac{dz}{z^2}\ ( q(z)+\bar q(z))
\right]\, .
\label{Q0}
\ee
The inverse transformation reads:
\be
q(x)+\bar q(x)= Q_0(x)+\frac{\sqrt x}{2} \int_x^1 \frac{dy}{y^{3/2}}\ Q_0(y)\, .
\ee

\item
The Mellin moments of functions $Q_{2\nu}(x,t)$ give the form factors of the
twist-2 local operators with fixed conformal spin, see details in Ref.~\cite{MaxAndrei}.
In particular, the $x$-moment of the lowest functions $Q_0(x,t)$ and $Q_2(x,t)$ gives the form factors
of the quark energy-momentum tensor:

\be
\nonumber
\int_0^1dx\ x\ Q_0(x,t)&=&\frac{5}{6}\ M_2^Q(t)\, , \\
\int_0^1dx\ x\ Q_2(x,t)&=&d_1(t)+\frac{5}{12}\ M_2^Q(t)\, .
\label{pravilasum}
\ee
Here $M_2^Q(t)$, $d_1(t)$ are form factors of the quark part of the energy momentum
tensor in notations of Ref.~\cite{sily}. $M_2^Q(0)$ is the fraction of the hadron momentum
carried by quarks, $d_1(t)$ is the leading coefficient in the Gegenbauer expansion of the D-term \cite{PW99}
and this form factor can be related to the spatial distribution of strong forces inside the hadron \cite{sily}.
In the case of spin-1/2 hadron the above sum rule
contains one additional form factor $J^Q(t)$, with $J^Q(0)$-the angular momentum in the
hadron carried by the quarks.

\item
The coefficients $d_n(t)$ of the Gegenbauer expansion of the D-term
\be
D(z,t)=(1-z^2)\sum_{n=1\atop \scriptstyle{\rm odd}}^{\infty}\ d_n(t)\ C_n^{(3/2)}(z)\, ,
\ee
can be computed with help of
the following generating function:
\be
\label{generatingd}
\sum_{n=1\atop \scriptstyle{\rm odd}}^{\infty}\ d_n(t)\ \alpha^n =
\frac{1}{\alpha}\ \int_0^1\ \frac{dz}{z}\ \sum_{\nu=0}^\infty
(\alpha\ z)^{2\nu} Q_{2\nu}(z,t) \left(\frac{1}{\sqrt{1+\alpha^2 z^2}}-\delta_{\nu 0}
\right)
\ee
\end{itemize}

\vspace{0.3cm}
\noindent
Now we give couple of new results concerning the dual parametrization.
Using the expression (\ref{integralrelation}) we can write the expansion of the GPD $H(x,\xi,t)$ around
the point $\xi=0$ with fixed $x$  ($x>0$) to the
order $\xi^2$ as follows:

\be
\nonumber
&&H(x,\xi,t)\sim \frac 12\ Q_0(x,t)+\frac{\sqrt x}{4} \int_x^1 \frac{dy}{y^{3/2}}\ Q_0(y,t)\\
\nonumber
&+&\frac{\xi^2}{8}\ \left[-\frac{1-x^2}{x}\frac{\partial}{\partial x} Q_0(x,t)+\frac{1}{8}\int_x^1\frac{dy}{y^3}\ Q_0(y,t)\left(3 \sqrt{\frac{y}{x}}-
\left(\frac{y}{x}\right)^{3/2}\right) \right.\\
\nonumber
&+&\left.
\frac{3}{8}\int_x^1\frac{dy}{y}\ Q_0(y,t)\left(
  \sqrt{\frac{x}{y}}+ \sqrt{\frac{y}{x}}\right) \right. \\
 \label{expansionH}
&+& \left.
\ Q_2(x,t)+\frac{3}{8}\int_x^1\frac{dy}{y}\ Q_2(y,t)\left(
\frac 12\ \sqrt{ \frac{x}{y}}+  \sqrt{\frac{y}{x}}+\frac {5}{2} \left(\frac{y}{x}\right)^{3/2}\right)\right]+O(\xi^4)\,.
\ee
It is very instructive expression, we see that the expansion of the GPD in small $\xi$ to the order $\xi^{2\nu}$ involves
only finite number of functions $Q_{2\mu}(x,t)$ with $\mu\leq \nu$ (to the order $\xi^2$ these are only $Q_0$ and $Q_2$).
If the expansion of GPD around $\xi=0$ is known (for example if the GPD is computed in a model), than it
can be used to determine the
functions $Q_{2\nu}(x,t)$ performing the small $\xi$
expansion order by order. This justifies the naming the functions $Q_{2\nu}(x,t)$--forward-like parton
distributions.

Another limiting case, when the Eq.~(\ref{kernel-podrobno}) simplifies is the limit $\xi\to 1$. In this limit
GPD has all properties of distribution amplitudes. This limit can be used to constrain functions $Q_{2\nu}(x,t)$
because sometimes one can derive the form of the GPD in this limit, for example pion GPDs are constrained
by the soft pion theorems \cite{MVP98}.

\be
\nonumber
H(x,\xi=1,t)&=&\frac 12 \ (1-x^2)\sum_{\nu=0}^\infty \int_0^1 dy\ \left[\frac{1}{(1-2 x y+y^2)^{3/2}}
-\frac{1}{(1+2 x y+y^2)^{3/2}}\right. \\
&-& \left.
\sum_{k=1\atop \scriptstyle{\rm odd}}^{2\nu-3} y^k\ C_k^{(3/2)}(x)\right]\ Q_{2\nu}(y,t)\, .
\ee
Expanding the integrand in the above equation in variable $y$ we obtain the Gegenbauer series
for $H(x,\xi=1,t)$:

\be
H(x,\xi=1,t)=(1-x^2)\sum_{k=1\atop \scriptstyle{\rm odd}}^{\infty} h_k(t)\ C_k^{(3/2)}(x)\, ,
\ee
with Gegenbauer coefficients computed as:
\be
h_k(t)=\int_0^1 dx\ x^k \sum_{\nu=0}^{\frac{k+1}{2}} Q_{2\nu}(x,t)\, .
\ee
In particular:

\be
h_1(t)=\int_0^1 dx\ x\  \left(Q_{0}(x,t)+Q_{2}(x,t)\right)=\frac 54\ M_2^Q(t)+d_1(t)\, ,
\ee
i.e. this coefficient is fixed completely in terms of form factors of the energy-momentum tensor. This, actually,
is not surprising as it is a consequence of the polynomiality condition.

Finally, we note that to ensure the finiteness of the D-term and existence of the Mellin moments of GPDs the small $x$
behaviour
of the function $Q_{2\nu}(x,t)\sim \frac{1}{x^\lambda}$ should be with $\lambda < 2\nu$ (for $\nu \ge 1$) and
$\lambda<2$ for $\nu=0$.

\section*{\normalsize \bf Forward-like distributions in terms of amplitudes: the inversion problem}
\noindent
The knowledge of GPD as a function of three variables is equivalent to the knowledge
of set of forward like parton distributions $Q_{2\nu}(x,t)$.
In this section we attempt to express the forward-like parton distributions
directly in terms of the amplitudes of hard exclusive processes.

The leading order amplitude of hard exclusive reactions is expressed
in terms of the following elementary amplitude\footnote{We restrict ourselves to the
singlet (even signature) amplitudes, generalization for odd signature amplitudes is trivial.}:
\be
A(\xi,t)=\int_{0}^1 dx\,H(x,\xi,t)\,
\left[\frac 1{\xi-x-i0} -\frac 1{\xi+x-i0}\right].
\label{elementaryAMP}
\ee
We see that the amplitude is given by the convolution integral in which dependence
of GPDs on variable $x$ is ``integrated out". We shall see below, that mathematically from the equations
(\ref{elementaryAMP}) one can not completely restore\footnote{We consider
the inversion problem at a single value of the large photon virtuality.}
the GPD $H(x,\xi,t)$. So we are not able to perform ``complete imaging" of the target hadron
from the knowledge of the amplitude.
The key question is: what part of the ``complete image" can be restored from the known amplitude?
What is the physics content of the restorable part of the complete image?
Below we address this question in terms of dual parametrization.

We can express the amplitudes in terms of forward-like functions
$Q_{2\nu}(x,t)$ as following \cite{MaxAndrei}:
\begin{eqnarray}
\nonumber
{\rm Im\ } A(\xi,t)&=&
\int_{\frac{1-\sqrt{1-\xi^2}}{\xi}}^1
\frac{dx}{x} N(x,t)\
\Biggl[
\frac{1}{\sqrt{\frac{2 x}{\xi}-x^2-1}}
\Biggr]\, ,\\
\nonumber
{\rm Re\ } A(\xi,t)&=&
\int_0^{\frac{1-\sqrt{1-\xi^2}}{\xi}}
\frac{dx}{x} N(x,t)\
\Biggl[
\frac{1}{\sqrt{1-\frac{2 x}{\xi}+x^2}} +
\frac{1}{\sqrt{1+\frac{2 x}{\xi}+x^2}}-\frac{2}{\sqrt{1+x^2}}
\Biggr]  \\
&+&\int^1_{\frac{1-\sqrt{1-\xi^2}}{\xi}}
\frac{dx}{x} N(x,t)\
\Biggl[
\frac{1}{\sqrt{1+\frac{2 x}{\xi}+x^2}}-\frac{2}{\sqrt{1+x^2}}
\Biggr]+ 2 D(t)
\, .
\label{REIM}
\end{eqnarray}
Here we introduced the function:
\be
N(x,t)=\sum_{\nu=0}^{\infty} x^{2\nu} Q_{2\nu}(x,t)\, ,
\label{Nifunction}
\ee
and the D-form factor:
\be
D(t)=\sum_{n=1}^\infty d_n(t)=\frac 12 \int_{-1}^1 dz\ \frac{D(z,t)}{1-z}\, .
\ee
Here $D(z,t)$ is the D-term \cite{PW99}.

Now we clearly see that the knowledge of  the LO amplitude is equivalent to
the knowledge of the function $N(x,t)$ and D-form factor $D(t)$. Moreover the D-form factor
can be computed in terms of $N(x,t)$ and $Q_0(x,t)$. Note that the latter function
is to great extend is fixed by the forward parton distributions, see Eq.~(\ref{Q0}). Indeed,
if we use Eq.~(\ref{generatingd}) at $\alpha=1$, we can write:

\be
D(t) =
 \int_0^1\ \frac{dz}{z}\
Q_{0}(z,t) \left(\frac{1}{\sqrt{1+ z^2}}-1
\right)+
 \int_0^1\ \frac{dz}{z}\
\left[N(z,t)-Q_{0}(z,t)\right] \ \frac{1}{\sqrt{1+ z^2}}
\label{DFFNQ0}
\ee
Note that in this equation all integrals are convergent if the functions $Q_{2\nu}(x,t)$ satisfy
the small $x$ behaviour discussed at the end of the second section. In Ref.~\cite{Teryaev}
the following representation for the D-form factor in terms of GPD has been suggested:

\be
D(t)=\int_{-1}^1\frac{dx}{x}\ \left[H(x,x,t)-H(x,0,t) \right]\ .
\ee
Unfortunately, this expression is divergent, as $H(x,x)$ and $H(x,0)$ have different coefficients in front
of leading small-$x$ asymptotic, see Refs.~\cite{Shuvaev:1999ce,MaxAndrei}.  The reason for this divergency is
that the above expression was obtained
by small $\xi$ expansion of GPDs, which explodes if $x\sim \xi$.

Another remarkable physics feature of the function $N(x,t)$ is that
its Mellin moments are related to the contributions of states with
fixed angular momentum in the $t$-channel:

\be
\int_0^1 dx\ x^{J-1}\ N(x,t)=\frac 12 \int_{-1}^1 dz \frac{\Phi_J(z,t)}{1-z}\, ,
\label{Jexchange}
\ee
where $\Phi_J(z,t)$ is the distribution amplitude corresponding to two quark exchange in the $t$-channel
with fixed angular momentum $J$. The quantity on RHS of Eq.~(\ref{Jexchange}) carries valuable information
about the hadron structure -- it tells how the target nucleon responses to the
well defined quark-antiquark probe of {\it arbitrary} spin $J$.

Now we turn to the inversion problem: how to obtain the function $N(x,t)$ if we know the amplitude of a hard exclusive
process. This problem is the central for the physics of hard exclusive processes. To solve the problem
we start with the expression (\ref{REIM}) for the imaginary part of the amplitude. It is useful to make
the following substitution for the integration variable $x$:
$$ \frac 1w = \frac 12\left(x+\frac 1x\right)\, .$$
This substitution correspond to famous Joukowski conformal map \cite{Zhu},
which historically was used to understand some principles of aerofoil design.
After this change of variables the expression for the imaginary part of the amplitude gets the form:

\be
{\rm Im\ } A(\xi,t)&=& \int_\xi^1 \frac{dw}{w}\ M(w,t)\ \frac{1}{\sqrt{\frac w \xi-1}}\, ,
\label{imsimple}
\ee
where the function $M(w,t)$ is related to the function $N(x,t)$ as:

\be
M(w,t)=N\left( \frac{1-\sqrt{1-w^2}}{w},t\right)\ \frac{w}{\sqrt{2(1-w^2)}\sqrt{1-\sqrt{1-w^2}}}\, .
\ee
Now we can easily invert Eq.~(\ref{imsimple}), this can be done in many ways. One of them is to
view the problem as the tomography problem and solve it with help of inverse Radon transformation \cite{Radon}.
The solution of the integral equation (\ref{imsimple}) is:
\be
M(w,t)=\frac{w}{\pi} \int_w^1 \frac{d\xi}{\xi^{3/2}}\ \frac{1}{\sqrt{\xi-w}}\ \left\{ \frac 12
{\rm Im\ } A(\xi,t)-\xi \frac{d}{d\xi}{\rm Im\ } A(\xi,t)
\right\}
\ee
Performing trivial change of variables we arrive at
the final result for the function $N(x,t)$:

\be
N(x,t)=\frac{2}{\pi}\ \frac{x(1-x^2)}{~~~(1+x^2)^{3/2}} \int_\frac{2 x}{1+x^2}^1\frac{d\xi}{\xi^{3/2}}\
\frac{1}{\sqrt{\xi-\frac{2 x}{1+x^2}}} \left\{ \frac 12{\rm Im\ } A(\xi,t)-\xi \frac{d}{d\xi}{\rm Im\ } A(\xi,t)
\right\}
\label{nina}
\ee
This remarkable formula allows to restore the function $N(x,t)$ from the measured imaginary part
of the amplitude. Note that the inversion formula contains the amplitude only in the physical region.
At $\xi\to 1$ the imaginary part of the amplitude should go to zero. Let us assume that
$
{\rm Im\ } A(\xi,t) \sim (1-\xi)^\beta
$ as $\xi\to 1$, than from Eq.~(\ref{nina}) one can easily obtain that

$$
N(x,t)\sim \frac{1}{2^{\beta-\frac 12}\pi} \frac{\beta}{\beta-1/2}\ (1-x)^{2 \beta}
$$
as $x\to 1$.

As to small $x$ behaviour, the Regge like asymptotic of the imaginary part of the amptlitude
${\rm Im\ } A(\xi,t) \sim 1/\xi^\alpha$, according to Eq.~(\ref{nina}), corresponds to the following
small $x$ behaviour of $N(x,t)$:

$$
N(x,t)\sim \frac{1}{2^\alpha}\ \frac{\Gamma(1+\alpha)}{\Gamma(\frac 12)\Gamma(\frac 12+\alpha)}\ \frac{1}{x^\alpha}\, .
$$
Note that the forward-like function $Q_{2\nu}(x,t)$ enter the function $N(x,t)$ with the weight $x^{2\nu}$. Given the
same small $x$ behaviour of these functions,
the contribution of higher functions to the small $\xi$ (large energies)
behaviour of the amplitude is suppressed by $\xi^{2 \nu}$. This resembles the contribution of daughter Regge trajectories
of dual resonance models (DRMs) for hadron interactions. The relations of the dual parametrization of GPDs to DRMs will
be discussed elsewhere.

Let us give also the expression for the Mellin moments of the function $N(x,t)$ which, as we discussed, are related to the
quark exchanges in the $t-$channel with fixed angular momentum, in terms of the amplitude. The corresponding
expression has the form:

\be
\int_0^1 dx x^{J-1}\ N(x,t)=\frac 1\pi\int_0^1\frac{d\xi}{\xi^{3/2}}\
 \left\{ \frac 12{\rm Im\ } A(\xi,t)-\xi \frac{d}{d\xi}{\rm Im\ } A(\xi,t)
\right\}\ R_J(\xi)\, ,
\ee
with the function $R_J(\xi)$ given by the following integral:

\be
R_J(\xi)=\int_0^\xi \frac{dw}{\sqrt w}\ \left(
\frac{1-\sqrt{1-w^2}}{w}\right)^{J+\frac 12}\ \frac{1}{\sqrt{\xi-w}}\, .
\ee
Its small $\xi$ behaviour has the form:
\be
R_J(\xi)\sim \left(\frac{\xi}{2} \right)^{J+\frac 12} \frac{\Gamma(J+1) \Gamma(\frac 12)}{\Gamma(J+\frac 32)}\, .
\ee

Finally, we can in principle compute the D-form factor with help of Eq.~(\ref{DFFNQ0}). For that calculation
we need additional information -- the function $Q_0(x,t)$. Fortunately this function can be fixed (up to t-dependence)
with help of Eq.~(\ref{Q0}). We do not give the complete expression
for the D-form factor, but rather the key integral in Eq.~(\ref{DFFNQ0}):

\be
\int_\epsilon^1 \frac{dx}{x}\ \frac{1}{\sqrt{1+x^2}}\ N(x,t)=\frac{\sqrt 2}{\pi}\int_{2\epsilon}^1
\frac{d\xi}{\xi}\ \left\{ \frac 12{\rm Im\ } A(\xi,t)-\xi \frac{d}{d\xi}{\rm Im\ } A(\xi,t)
\right\}\, ,
\ee
where we introduced infinitesimally small regulating parameter $\epsilon$ to cut possible divergencies at small
$x$. In principle, from the measurements of the imaginary part of the amplitude
we can determine the function $N(x,t)$ and from measurements of the real part the D-form factor $D(t)$. These two
measurements would allow to constrain considerably the function $Q_0(x,t)$ which gives directly the probability
density of partons in 3D space \cite{Bur}.

To summarize this section:
all above formulae give us the possibility access the
wide set of hadrons properties directly in terms of amplitudes
of the hard exclusive processes. Still we are not able to restore the complete GPD as for this we need to determine
not only the function $N(x,t)$ but rather the whole set of functions $Q_{2\nu}(x,t)$. This can be achieved
if we would know more general function:

\be
N_\alpha(x,t)=\sum_{\nu=0}^\infty (\alpha x)^{2\nu}\ Q_{2\nu}(x,t).
\label{Nalpha}
\ee
In principle, this function can be determined if we would measure the logarithmic scaling violation
for the amplitude. Probably, in practice, it is difficult problem and more practical way
is to model the function $N_\alpha(x,t)$ using the non-perturbative information about hadron structure.
We shall discuss possible ways for such modelling after we derive dispersion relations and
crossing in the next section.

\section*{\normalsize \bf Dispersion relations and crossing}
\noindent
Up to now we made use of only the imaginary part of the amplitude, let us study the real part of the amplitude.
For this we take the expression (\ref{nina}) for the $N(x,t)$ in terms of ${\rm Im} A(\xi,t)$ and substitute
it into Eq.~(\ref{REIM}). After some simple calculations we arrive to the following expression for real part
of the amplitude:

\be
{\rm Re} A(\xi,t)= 2 D(t)+\frac{1}{\pi}\ vp \int_0^1 d\zeta \ {\rm Im} A(\zeta,t) \left(\frac{1}{\xi-\zeta}-
\frac{1}{\xi+\zeta} \right)\, ,
\label{DR}
\ee
in which we can immediately recognize the dispersion relation for the amplitude with one subtraction
at non-physical point $\xi=\infty$ (corresponding to $\nu=(s-u)/4m=0$). The D-form factor is the corresponding
subtraction constant. This result was obtained recently in Refs.~\cite{Teryaev,DiehlIvanov} by independent methods.
We see that the dual parametrization automatically ensures the dispersion relations for the amplitudes.

The very idea of the dual parametrization of GPDs in terms of $t$-channel exchanges was motivated by the crossing
relations
\cite{MVP98} between GPDs and generalized distribution amplitude \cite{DiehlTeryaev}. The later enter the description
of the hard exclusive processes in the cross channel, like $\gamma^*+\gamma\to h+\bar h$. In the LO
the amplitude of the cross process can be expressed in terms of the function $N(x,t)$ analytically continued
to time-like $t$ ($t>0$)\footnote{Such continuation can be performed with help of dispersion relations in variable
$t$, see e.g. Ref.~\cite{MVP98}}:

\be
A^{\rm cross}(\eta,t)=
\int_0^{1}
\frac{dx}{x} N(x,t)\
\Biggl[
\frac{1}{\sqrt{1-2 x \eta+x^2}} +
\frac{1}{\sqrt{1+ 2 x\eta+x^2}}-\frac{2}{\sqrt{1+x^2}}
\Biggr]  + 2 D(t)\, .
\ee
Here $-\eta$ is directly related  to the $\cos(\theta_{cm})$-- cosine of scattering angle in centre of mass system,
see for details Refs.~\cite{DiehlTeryaev,MVP98}. Now substituting our inversion formula (\ref{nina}) into this expression
we obtain, rather simple result:

\be
A^{\rm cross}(\eta,t)=\frac 2\pi \int_0^{|\eta|} d\xi \frac{\xi}{1-\xi^2}\ {\rm Im} A\left(\frac {\xi}{|\eta|},t\right)
+2\ D(t)\, .
\ee
Actually this equation is the consequence of the dispersion relation (\ref{DR}).

\section*{\normalsize \bf A way to model $Q_{2\nu}(x,t)$}
\noindent
As we mentioned above the complete knowledge of GPD is equivalent to the knowledge of
function (\ref{Nalpha}). In this section we consider possible ways to model this function. For
simplicity we do not consider the $t-$ dependence and that why we do not write the corresponding
argument.
Without loss of generality we can represent the function $Q_{2\nu}$ as the Mellin convolution:
\be
Q_{2\nu}(x)=\int_x^1 \frac{dz}{z}\ Q_0(z)\ P_\nu\left(\frac xz \right)\, ,
\label{mellinconv}
\ee
then the function $N_\alpha(x)$ can be written as the integral transform of the $Q_0(x)$:

\be
N_\alpha(x)=\int_x^1 \frac{dz}{z}\ K\left(\alpha x,\frac xz \right) \ Q_0(z)\, ,
\ee
where the integral kernel is:

\be
K(x,y)=\sum_{\nu=0} x^{2\nu} P_\nu(y)\, .
\label{Kkernel}
\ee
The presentation of $Q_{2\nu}$ as the Mellin convolution (\ref{mellinconv})
has an advantage as the functions $P_\nu(y)$ and hence the integral kernel (\ref{Kkernel}) are
scale independent at LO. It is little known about the kernel (\ref{Kkernel}), its small $x$ expansion has the
form:

\be
K(x,y)\sim \delta(1-y)+x^2\ P_1(y)+\ldots \, .
\ee
We make the following ansatz for the integral kernel $K(x,y)$:
\be
K(x,y)=(1+A(x))\delta(1-y)+ B(x)\, .
\label{ansatz}
\ee
Surely, this ansatz is very simple, but still it gives enough freedom in modelling of the function $N_\alpha(x,t)$.
Let us assume that apart from the knowledge of the amplitude we have an additional non-perturbative information.
For instance, one possess the knowledge of the D-term. Using the generating function (\ref{generatingd}) we easily obtain
its relation to functions $A(x)$ and $B(x)$ in our ansatz (\ref{ansatz}):

\be
\sum_{n=1}^\infty \alpha^n\ d_n=\frac 1\alpha
\int_0^1 \frac{dx}{x}\ Q_0(x)\ \left\{
\frac{1+A(\alpha x)}{\sqrt{1+\alpha^2 x^2}}-1 +\int_0^x \frac{dz}{z} \frac{B(\alpha z)}{\sqrt{1+\alpha^2 z^2}}
\right\}\, .
\ee
If the functions $A(x)$ and $B(x)$ are related to each other by the following equation:

\be
(1+x^2) (x A'(x)+B(x))=x^2(1+A(x))\, ,
\label{null}
\ee
the corresponding D-term is zero.
This relation can be used for modelling of the GPDs for which the D-term is identically zero, e.g.
the combination $H(x,\xi)+E(x,\xi)$ of the nucleon GPDs.  For general case we can use the following
ansatz:

\be
A(x)=C(x)+A^{(0)}(x),\ \ \ B(x)=B^{(0)}(x),
\ee
where $A^{(0)}(x)$ and $B^{(0)}(x)$ satisfy Eq.~(\ref{null}). The function $C(x)$ can be determined from known
coefficients $d_n$
and known function $Q_0(x)$ through the following relation:
\be
\sum_{n=1}^\infty \alpha^n\ d_n=\frac 1\alpha
\int_0^1 \frac{dx}{x}\ Q_0(x)\ \left(
\frac{1+C(\alpha x)}{\sqrt{1+\alpha^2 x^2}}-1\right)
\ee

If we look for the solution of Eq.~(\ref{null}) in class of polynomials of finite order $2 N$, the
resulting polynomials $A^{(0)}(x)$ and $B^{(0)}(x)$ depend on $N-1$ free parameters. For example the solution
in the space of the 4th order polynomials depends on one free parameter $a$ and has the form:

\be
A^{(0)}(x)=a x^2 + (1- a) x^4, \ \ \ B^{(0)}(x)= (1- 2 a) x^2 - 3(1-a) x^4 \, .
\ee
The expression for the function $N_\alpha(x)$ in this case has the form:

\be
\nonumber
N_\alpha^{(0)}(x)&=&[1+a\ ( \alpha x)^2 + (1- a)\ ( \alpha x)^4]\ Q_0(x)
\\
&+&
[(1- 2 a)\ (\alpha x)^2 - 3(1-a)\ ( \alpha x)^4] \int_x^1\frac{dz}{z}\ Q_0(z)\, .
\ee
Allowing solutions in the class of higher order polynomials we introduce new free parameters which can be fitted to
presumably known
function $N(x)$, the latter can be obtained from the data with help of Eq.~(\ref{nina}).

\section*{\normalsize\bf Conclusions}
In the framework of dual parametrization of GPDs we derive explicit equation (\ref{nina})
which expresses particular combination  (\ref{Nifunction}) of forward-like functions $ Q_{2\nu}(x,t)$
in terms of the imaginary part of the process amplitude. The real part of the amplitude contains
one additional constant -- the D-form factor, which is actually the subtraction constant of the corresponding
dispersion relation. We show that the D- form factor can be computed in terms of the function $N(x,t)$
and the forward-like function $Q_0(x,t)$ which is at $t=0$ is related to forward parton distributions
by Eq.~(\ref{Q0}). It means that the amplitudes of hard exclusive processes (at fixed hard scale)
provide us only with partial image of total GPDs. This partial image is usefully encoded in the
function $N(x,t)$ and D-form factor $D(t)$ [equivalently in $N(x,t)$ and forward-like function $Q_0(x,t)$].
The function $N(x,t)$ contains direct information about the two quark exchange amplitude in t-channel with
fixed angular momentum. We showed that the $J$'s Mellin moments of $N(x,t)$ gives the response of the target
hadron to the (string) quark-antiquark operator with spin $J$.

\section*{\normalsize\bf Acknowledgements}
We are thankful to  N.~Kivel and D.~M\"uller for many
valuable discussions. The
work is supported
by the Sofja Kovalevskaja Programme of the Alexander von Humboldt
Foundation, the Federal Ministry of Education and Research and the
Programme for Investment in the Future of German Government.


\begin{thebibliography}{99}
\bibitem{pioneers}
D. M\"uller, D. Robaschik, B. Geyer, F.M. Dittes, and J. Horejsi,
Fortschr.~Phys. {\bf 42}, 101 (1994);\\
X.~D.~Ji,
  Phys.\ Rev.\ Lett.\  {\bf 78} (1997) 610
  [arXiv:hep-ph/9603249];\\
   A.~V.~Radyushkin,
  Phys.\ Lett.\  B {\bf 380} (1996) 417
  [arXiv:hep-ph/9604317];\\
X.~D.~Ji,
  Phys.\ Rev.\  D {\bf 55} (1997) 7114
  [arXiv:hep-ph/9609381].

\bibitem{GPV}
K.~Goeke, M.~V.~Polyakov and M.~Vanderhaeghen,
Prog.\ Part.\ Nucl.\ Phys.\  {\bf 47} (2001) 401
[arXiv:hep-ph/0106012].
\bibitem{Diehlrev}
  M.~Diehl,
  Phys.\ Rept.\  {\bf 388} (2003) 41
  [arXiv:hep-ph/0307382].

\bibitem{Belitskyrev}
  A.~V.~Belitsky and A.~V.~Radyushkin,
  Phys.\ Rept.\  {\bf 418} (2005) 1
  [arXiv:hep-ph/0504030].

\bibitem{Radon}
J.~Radon, Berichte S\"achsische Akademie
der Wissenschaften, Leipzig, Math-Phys. Kl., {\bf 69} (1917) 262;\\
 S.~Deans, {\it The Radon transform and some
of its applications}, Wiley-Interscience, 1983.

\bibitem{Teryaev:2001qm}
  O.~V.~Teryaev,
  Phys.\ Lett.\  B {\bf 510}, 125 (2001)
  [arXiv:hep-ph/0102303].



\bibitem{Mueller:2006pm}
  D.~Mueller,
  arXiv:hep-ph/0605013.
\bibitem{Kumericki:2007sa}
  K.~Kumericki, D.~Muller and K.~Passek-Kumericki,
  arXiv:hep-ph/0703179.



\bibitem{MaxAndrei}
  M.~V.~Polyakov and A.~G.~Shuvaev,
  arXiv:hep-ph/0207153.

\bibitem{MVP98}
M.V.~Polyakov,
Nucl.~Phys. {\bf B555}, 231 (1999).
\bibitem{Guzey}
  V.~Guzey and M.~V.~Polyakov,
  Eur.\ Phys.\ J.\  C {\bf 46}, 151 (2006)
  [arXiv:hep-ph/0507183];\\
 V.~Guzey and T.~Teckentrup,
  Phys.\ Rev.\  D {\bf 74}, 054027 (2006)
  [arXiv:hep-ph/0607099].


\bibitem{PW99}
M.V. Polyakov and C. Weiss, Phys.~Rev.~D {\bf 60}, 114017 (1999).


\bibitem{sily}
  M.~V.~Polyakov,
  Phys.\ Lett.\  B {\bf 555}, 57 (2003)
  [arXiv:hep-ph/0210165].

\bibitem{Shuvaev:1999ce}
  A.~G.~Shuvaev, K.~J.~Golec-Biernat, A.~D.~Martin and M.~G.~Ryskin,
  Phys.\ Rev.\  D {\bf 60} (1999) 014015
  [arXiv:hep-ph/9902410].

\bibitem{Zhu}
N.E. Joukovskii, {\it On the problem of cutting vortex filaments}, Mat. Sbor. {\bf 17} (1895) No 4
702–719 (in Russian);\\
N.E. Joukovskii, {\em De la chute dans l´air de corps l\'{e}gers de forme allong\'{e}e, anim\'{e}s d´un mouvement
rotatoire}, Bull. Inst. A\'{e}rodyn. Koutchino (1906) No 1 51–65.

\bibitem{Teryaev}
 O.~V.~Teryaev,
  arXiv:hep-ph/0510031;\\
  I.~V.~Anikin and O.~V.~Teryaev,
  arXiv:0704.2185 [hep-ph].

\bibitem{DiehlIvanov}
  M.~Diehl and D.~Y.~Ivanov,
  arXiv:0707.0351 [hep-ph].



\bibitem{DiehlTeryaev}
M.~Diehl, T.~Gousset, B.~Pire and O.~Teryaev,
Phys.\ Rev.\ Lett.\  {\bf 81} (1998) 1782
[arXiv:hep-ph/9805380].

\bibitem{Bur}
M.~Burkardt,
Phys.\ Rev.\ D {\bf 62} (2000) 071503
[arXiv:hep-ph/0005108].

M.~Burkardt,
arXiv:hep-ph/0207047.



\end{thebibliography}
\end{document}